\documentclass[journal=jacsat,manuscript=article]{achemso}

\usepackage[version=3]{mhchem} % Formula subscripts using \ce{}
\usepackage[T1]{fontenc}       % Use modern font encodings
\usepackage{graphics}
\usepackage[usenames, dvipsnames]{color}
\usepackage{longtable}
\usepackage{multirow}
\usepackage{makecell}
\usepackage{natmove}
\setkeys{acs}{usetitle = true}
\usepackage{amsmath,achemso}
\usepackage{bm}
\setkeys{acs}{maxauthors = 9}

\author{Ngoc Thanh Thuy Tran}
\affiliation {Hierarchical Green-Energy Materials (Hi-GEM) Research Center,
National Cheng Kung University, Tainan 70101, Taiwan}
\email{tranntt@phys.ncku.edu.tw}
\author{Godfrey Gumbs}
\affiliation{Department of Physics and Astronomy, Hunter College of the City University of New York, New York, USA}
\author{Duy Khanh Nguyen}
\affiliation{Laboratory of Applied Physics, Advanced Institute of Materials Science, Ton Duc Thang University, Ho Chi Minh City, Vietnam and Faculty of Applied Science, Ton Duc Thang University, Ho Chi Minh City, Vietnam}
\author{Ming-Fa Lin}
\affiliation {Department of Physics, National Cheng Kung University, Tainan, Taiwan}
\title[An \textsf{achemso} demo]
  {Fundamental Properties of Metal-Adsorbed Silicene: A DFT Study}


\begin{document}

%%%%%%%%%%%%%%%%%%%%%%%%%%%%%%%%%%%%%%%%%%%%%%%%%%%%%%%%%%%

\begin{abstract}

Sodium, magnesium and aluminum adatoms, which, respectively, possess one, two and three valence electrons in terms of 3s, 3s\(^2\), and (3s\(^2\), 3p) orbitals, are very suitable for helping us understand the adsorption-induced diverse phenomena. In this study, the revealing properties of metal (Na/Mg/Al)-adsorbed graphene systems are investigated by mean of the first-principles method. The single- and double-side chemisorption cases, the various adatom concentrations, the hollow/top/valley/bridge sites, and the buckled structures are taken into account. The hollow and valley adsorptions, which, respectively, correspond to the Na/Mg and Al cases, create the extremely non-uniform environments within the Moire superlattices. This lead to diverse orbital hybridizations in Na/Mg/Al-Si bonds, as indicated from the Na/Mg/Al-dominated bands, the spatial charge density distributions and the orbital-projected density of states (DOS). Among three kinds of metal-adatom adsorptions, the Al-adsorption configurations present the  strongest chemical modifications. The ferromagnetic configurations are shown to only survive in the specific Mg- and Al-adsorptions, but not the Na-cases. The theoretical predictions could be validated by experimental measurements and the up-to-date potential applications are included. Furthermore, the important similarities and differences with the graphene-related systems are also discussed.

\end{abstract}\
\section{I. Introduction}
 Since graphene, the first two-dimensional (2D) material, was successfully fabricated, the feverish scientific activity relating to graphene still keep unabated. Significant efforts on exploiting other graphene-related materials are continued reported for expanding potential applications.\cite{1,2} Following graphene, honeycomb lattice of silicon atoms namely silicene has recently received an increased interest.\cite{3,4,5} It exhibits a lot of remarkable properties arising from the hexagonal symmetry and Dirac-cone structure, e.g., the high charge carriers mobility,\cite{4} excellent optical absorption.\cite{5}  Different from the flat structure of graphene with sp$^2$ hybridization of carbon atoms, silicene posseses low-buckle as a result of the mixed sp$^2$-sp$^3$ hybridizations of silicon atoms.\cite{3} This make silicene surpassed graphene in some aspects, including greater spin-orbit coupling (SOC) and quantum spin Hall effect,\cite{6} easier tunable bandgap,\cite{7} stronger valley polarization and other related properties.\cite{8} These remarkable characteristics make silicene become potential materials in various application fields such as field effect transistors (FET),\cite{9,10} Li/Na-batteries,\cite{11,12} spintronic and valleytronic devices,\cite{13} sensor,\cite{14} etc.
 
 Up to now, single-layer silicene has been successfully synthesized by various experimental methods. The main method is bottom-up approach as so-called epitaxial growth on a substrate, whether by deposition onto a supporting template e.g. Ag(111)\cite{15}, Ir(111)\cite{16} or by segregation on a buffer layer e.g. ZrB$_2$.\cite{17} To further enhance the applications of silicene, it is highly desirable to grow silicene on non-metallic substrate such as MoS$_2$,\cite{18} SiC.\cite{19} The graphene-like honeycomb lattice of silicene on metal substrates has been observed by the scanning tunneling microscope (STM).\cite{31} Due to the buckle structure, silicene is a direct-gap semiconductor with a negligible gap.\cite{3} The free-carrier density enhancement or the gap opening is considered as an important issue for improving the promissing applications of silicenes. The electronic properties can be easily modulated by dopings,\cite{20,21,22,23,24} external electric and magnetic fields,\cite{25,26} and mechanical strains.\cite{27,28} Among these modulations, chemical modification on the silicene surface is the most effective one in creating essential changes between the semiconducting and metallic behaviors (the non-magnetic and magnetic configurations). The chemical adsorption of metal adatoms is predicted to induce metallic band structures with high conduction electron densities.\cite{29,30} However, the relations among the geometric structure, electronic properties, and magnetic configuration towards various adatom coverages have not been thoroughly reported. The single/multi-orbital hybridizations between adatoms and silicene, a key point in understanding the electronic properties modifications, is only slightly mentioned in several studies. Sodium, magnesium, and aluminum, respectively, possess one, two, and three outermost electron orbitals, which are expected to be capable of providing extra carrier density and sufficient metallization for silicene system. Here, Na-, Mg- and Al-doped silicene materials will be able to create various orbital hybridizations, diversify the fundamental properties, and provide more information regarding potential applications. 

The first-principles calculations on Na, Mg and Al-adsorbed silicene systems include determining the metallization phenomena, 2D free carrier density, the low-lying extended states \(\pi\)-bonding on a honeycomb lattice, the well-defined/modified/thoroughly suppressed Dirac cone, the rigid shift \(\sigma\)-bands/the undefined ones, and the spin-degenerate or spin-split energy. They are expected to be greatly diversified by the different single-/multi-orbital hybridizations in Na-/Mg- and Al-Si bonds. How many orbitals of metal adatoms and silicon host atoms take part in the significant chemical bondings will be resolved in this work. The single- or multi-orbital hybridizations are clearly identified from the atom dominance of energy band, the orbital-dependent charge distribution, and the orbital-projected DOSs. They are responsible for the optimal geometrical structure, the unusual electronic properties, and can even diversify magnetic behaviors. Whether the ferromagnetic spin configuration could survive under the specific adatoms and adsorption configurations is thoroughly examined by the use of total magnetic moments, spin-split energy bands, spin-density distributions and spin-decomposed DOSs. The above-mentioned results can be examined by experimental measurements, such as angle resolved photoemission spectroscopy (ARPES),\cite{32} and scanning tunneling spectroscopy (STS).\cite{33} Our reliable and complete results should be very useful in the design and development of promissing device applications.

\section{II. Computational methods}

We note here that our first-principles calculations were performed with the use of the density functional theory (DFT) based Vienna ab initio simulation software package (VASP).\cite{34,35} The Perdew-Burke-Ernzerh of generalized gradient approximation method\cite{36} was employed to evaluate  the electron-electron Coulomb interactions while the electron-ion interactions belong to the projector augmented wave method.\cite{37} The spin configurations are taken into account to meticulous explore the chemical adsorption effects on the magnetic properties. In order to suppress the van der Waals interactions between two neighboring cells, the vacuum distance along the z-axis is set to be \( 15\)  \AA. A plane-wave basis set, with a maximum energy cut off of 500 eV is available in the calculations of Bloch wave functions. All atomic coordinates are relaxed until the change of eigenvalues between two simulation steps is less than 10$^{-5}$ eV, and the Hellmann-Feynman force convergence on each atom is set to be \(0.01  \) eV/\AA. The pristine first Brillouin zone is sampledby  \( 30 \times 30 \times 1 \) and  \( 100 \times 100 \times 1 \) k-points within the Gamma scheme for structure relaxations and further evaluations on electronic properties, respectively. Equivalent k-point mesh is built for other enlarged cells depending on their sizes. Furthermore, the van der Waals force, which utilizes the semiemprical DFT-D2 correction of Grimme\cite{38} is very useful in understanding the significant atomic interactions between silicene and adatom layers at high doping concentrations.

\section{III. Results and discussions}

Sodium, magnesium and aluminum adatom adsorption on monolayer silicene surfaces presents diversified geometric structures, mainly due to the different interlayer chemical bondings. Sodium and magnesium adatoms prefer to be adsorbed on the hollow site of silicene, while aluminum adatoms are more stable at the valley site (Fig. 1). This could be verified using the STM measurement.  Since the valley and top positions possess different chemical environments; they are capable of creating diversified essential properties.

\subsection{Na adsorptions}

Sodium-adsorbed silicene systems are stable under double- and single-side absorption, with specific concentrations and distributions (Fig. S1). The highest saturation can reach 100\% for double-sided adsorption. In the case of pristine silicene, the optimized Si-Si bond length and buckling are 2.28 \AA and 0.48 \AA, respectively, which are consistence with previous studies.\cite{38a,38b} At sufficiently low Na-concentration \( <12.5 \%\) the degree of buckling is comparable with that of pristine system, while it might be greatly enhanced for higher concentrations. That the \(\pi\) and \(\sigma\) bondings are non-orthogonal to each other becomes more obvious with an increase of Na concentration. The Na-Si bond lengths lie in the range of ${\sim\,3.10}$ \AA \ - ${3.26}$ \AA \, weakly dependent on the adsorption cases (Table 1). The nearest Si-Si bonds are slightly stretched by the observable change from 2.28 \AA \ to ${\sim\,2.30}$ \AA \ - ${2.33}$ \AA . This indicates the partial charge transfer through the interlayer chemical bonding. According to the above-mentioned geometric features, the \(\sigma\)-electronic chemical bondings, which arise from the (3s, 3p\(_x\), 3p\(_y\)) orbitals of silicon atoms, are deduced to hardly take part in the alkali-adatom chemisorption. As a consequence, the single-orbital hybridization of Na-3s and Si-3p\(_z\) will dominate the adsorption-induced chemical bonding and thus the other essential properties. 

The pristine silicene (Fig. 2(a)) exhibits a Dirac cone structure as marked by the red cirle at the K point with a negligible energy gap around 0.5 meV. This gap could be enlarged to 1.55 meV under the effect of spin-orbit coupling.\cite{6} These energy bands are predominantly dominated by 3p\(_z\)  orbitals of the nearest Si atoms as so-called $\pi$ bands and turn into parabolic dispersions at saddle M point. The unoccupied and occupied $\sigma$ parabolic bands are respectively initiated at about $\pm$ 1.2 eV. Compared with pristine silicene, the sodium-adsorbed systems exhibit unusual valence and conduction bands (Figs. 2(b)-2(h)). The main features of band structures include the highly asymmetric energy spectra about the Fermi level ($E_F$), the modified Dirac-cone structure with/without an energy spacing, the blue shift of $E_F$ relative to the Dirac point (the metallic behavior), the roughly rigid \(\sigma\) valence bands, the splitting and anti-crossing states, the Na- or (Na, Si)-dominated conduction/valence states, and the extra critical points in the energy-wave-vector space related to band hybridizations. The anisotropic and distorted Dirac cone which mainly arises from the 3p\(_z\) orbitals are initiated from the stable \(\Gamma /K\) valley (Fig. S2) as a result of the zone-folding effect. The initial/bottom valence one is very close to the first \(\sigma\) energy subbands with the concave-downward dispersion relations, where an energy spacing of \(~0.20\) eV corresponds to the deeper state. On the other hand, the sodium adsorptions on silicene surface are able to create the free conduction electrons with the sufficiently high carrier densities as the so-called n-type doping phenomena. Apparently, the Fermi level is dramatically changed from the middle of the well-behaved Dirac-cone structure (Fig. 2(a)) into the conduction cone of the modified one. That is to say, only the partial 3s orbitals in sodium adatoms serve as the conduction electrons. The Na-induced free electrons correspond to the occupied states between \(E_F\) and conduction Dirac point. The Na-adatoms only make minor contributions to the deeper valence bands. In general, there exist the separation of valence and conduction Dirac points, being roughly attributed to the different site energies (the ionization ones) of Na-3s and Si-3p\(_z\) orbitals. Specifically, the spin-split electronic structures are absent for the various Na-adsorption cases, i.e., there are no spin-dependent ferromagnetic/anti-ferromagnetic configurations.

The atom-, orbital- and spin-projected van Hove singularities density of states (DOS), as clearly illustrated in Figs. 3(a)-3(h), are available in identifying the critical orbital hybridizations Na-adsorbed silicenes and thus the complicated energy spectra and the existences of \(\pi\) and \(\sigma\) bondings. The 3s-3p\(_z\) orbital hybridization in the Na-Si bond is able to create the unusual van Hove singularities in DOS. This result strongly depends on Na-concentration, and it becomes vanishing at the sufficiently low concentrations \( <3.1 \%\) in Fig. 3(h). The Dirac-cone-induced structure has an obvious red shift relative to \(E_F\), clearly illustrating the n-type doping behavior. The covered area between the bottom of conduction dip and the Fermi level determines the Na-created free electron density. Roughly speaking, the valence and conduction dip is dominated by the Si-3p\(_z\) orbitals (the red curves). However, the Na-3s orbitals (the cyan curves) also make certain contributions to the latter at very high adatom concentrations \( >50 \%\) in Figs. 3(b) and 3(c), mainly owing to Na-Na interactions. That both 3p\(_z\) and 3s orbitals possess the merged van Hove singularities clearly indicates the significant interlayer single-orbital chemical bondings. There exist the splitting \(\pi\) and \(\pi ^*\) prominent peaks, being accompanied with extra minor structures, are attributed to the zone folding, the interlayer atomic interactions, and the enhanced buckling.

As to the \(\sigma\) electrons of silicon host atoms, the 3p\(_x\) and 3p\(_y\) orbitals (the pink and dashed blue curves) display the identical/different contributions in the presence of a symmetric/asymmetric environment under the \textbf{xy}-plane projection (Fig. S1). The initial \(\sigma\)-electronic valence band create the obvious shoulder structures as marked by the green arrow in the range -1.65 eV\({\rm  <  E  < }\)-1.20 eV, depending on Na-concentrations, being deeper than the pristine case (Fig. 3(a)). The whole band width covers the first shoulder, the symmetric peak, and the second one, which correspond to the parabolic, saddle and parabolic band-edge states at the \(\Gamma \), \({\rm M}\) and \({\rm K}\) points (Figs. 2(a)- 2(h)), respectively. Moreover, few van Hove singularities, which are associated with the four orbitals of Si atoms, occur simultaneously (the red, pink, dashed blue and green curves), e.g., those at -4.0 eV\({\rm  <  E  < }\)-3.20 eV related to the anti-crossings of \(\pi\) and \(\sigma\) bands. In short, the modified \(\pi\) and \(\sigma\) chemical bondings could be approximately identified from the specific van Hove singularities, and such examinations are supported by the band structures and spatial charge densities.   

The spatial charge distributions and their variations (\(\rho \) and \(\Delta\rho \)) after sodium adsorptions are capable of examining the changes of \(\pi\) and \(\sigma\) chemical bondings in silicene surface and the coupling-induced orbital hybridizations, as clearly illustrated in Fig. 4. The similar bonding phenomena could be revealed in other alkali-adsorbed silicene systems. Apparently, two kinds of orbital bondings in honeycomb lattice could survive simultaneously under any sodium-adsorption cases. The \(\sigma\)-electronic orbital hybridizations, with a very high charge density between two Si atoms, are hardly affected by the sodium adsorptions, corresponding to the almost vanishing charge density difference \(\Delta\rho \). This is responsible for the rigid red-shift of \(\sigma\) energy bands and the absence of valence bands co-dominated by Na and Si atoms (Figs. 2(a)-2(f)). On the other hand, the significant modifications on the \(\pi\) bondings are observable through the charge variations between Na and Si atoms (the heavy red rectangles), in which part of the adatom electrons is transferred to the host-atom region (the black arrow). Furthermore, they are relatively easily observed under the high-concentration condition (Fig. 4(c)). These features suggest the existence of only a single 3s-3p\(_z\) orbital hybridization in the Na-Si bonds, accounting for the (Na, Si)-co-dominated certain conduction bands. Specifically, the observable charge variation on \textbf{yz}-plane is revealed between two neighboring Na atoms at the sufficiently high concentrations, mainly owing to the creation of Na-Na bonds. 

\subsection{Mg adatoms}

While the structures are similar to those of Na-adsorptions, the Mg-Si bond lengths/the adatom heights (${\sim\,2.7}$ \AA \ /1.65 \AA ) are much shorter than the Na-Si ones (${\sim\,3.15}$ \AA). The \(100\%\) Mg-adsorption also could survive under the double-side adsorption case (Table 2), in which there exist the sufficiently strong Mg-Mg bonds. Mg-adsorption has the largest buckling of honeycomb lattice among three kinds of metal-adatom adsorptions at high concentrations \( >33 \%\). This enhanced buckling is expected to induce more important sp\(^3\) bondings (non-orthogonality of \(\pi\) and \(\sigma\) bondings) and spin-orbital couplings in silicene. These geometric features suggest the sufficiently strong chemical bondings between Mg adatoms and Si atoms, being associated with two 3s valence orbitals of Mg. The \(\pi\) bonding, which is extended on the silicene surface, will be strongly modified or even thoroughly disappear through the critical interlayer atomic interactions.

The Mg-adsorptions are able to induce the unusual band structures, as clearly indicated in Figs. 5(a)-5(d). All of them are belong 2D gapless metals/semimetals, in which the Mg adatoms create the free carriers. Roughly speaking, there exist the drastically modified low-lying energy dispersions, mainly owing to the extremely non-uniform environment in an enlarged unit cell and more important/complicated interlayer orbital hybridizations. The unusual bands are mostly contributed by the seriously distorted \(\pi\) bonding in silicene surface. Furthermore, the initial \(\sigma\) valence energy subbands come to exist at the \(\Gamma \) valley using the concave-downward dispersions in the range of \(E^v  <  - 1.20\) eV. Very interesting, the Mg-/(Mg,Si)-induced energy bands present the partial flat or oscillatory dispersions, being very close to, or even being merged with the low-lying \(\pi\)/\(\sigma\)-electronic structures. This phenomenon could survive under any Mg-adsorption cases. Since the Mg adatoms possess two 3s orbitals, they are able to create the Mg-/(Mg, Si)-dominated conduction and valence states. The former phenomena are very obvious at the sufficiently high concentrations, such as the \(50\%\) Mg-doping cases (Figs. 5(a)). However, with the low concentrations (Figs. 5(c) and 5(d)), the Mg-induced energy bands are dominated near the Fermi level.

On the other hand, the ferromagnetic configurations, which are clearly characterized by the spin-split valence and conduction energy bands and the net magnetic moments (Fig. 5 and Table 2), are further examined from the spin density distributions of the specific Mg-adsorption cases. For example, the \(16.7\%\) and \(3.1\%\) configurations as clearly displayed in Figs. 5(b) and 5(d) exhibit the unusual ferromagnetic phenomena. Apparently, the Mg adatoms of are able to create the dominating spin density associated with two 3s-orbital electrons of Mg adatoms while the host Si atoms make a minor contribution. They are able to induce the on-site Coulomb interactions in the intrinsic Hamiltonian.\cite{39} As a result, the orbital hybridizations, the spin-dependent many-particle interactions and spin-orbital couplings need to be included in the Hubbard tight-binding model.\cite{40}  

All the Mg-adsorbed silicenes under any cases, as clearly displayed in Figs. 6(a)-6(d), present a strongly/moderate/slightly modified dip with an observable/very narrow/negligible energy spacing below the Fermi level. The DOS is finite at the Fermi level for all the Mg-adsorption cases, in which its magnitude is very large for the partially flat bands/the oscillatory energy dispersions. The bottom of conduction part presents a red shift relative to \(E_F\). Furthermore, the special structures the red curves (due to the 3p\(_z\) orbitals) near Fermi level which mainly arise from the \(\pi\)- and \(\pi^*\)-electronic states are merged together. Very interesting, the strongly merged special structures are revealed in the Si-3p\(_z\) and Mg-3s orbitals (the red and cyan curves), where the certain energy regions cover the valence and conduction states. Specifically, the Mg-3s orbitals have an obvious valence- and conduction-state region of -2.0 eV\({\rm  <  E  < }\)2.0 eV under a very high concentration (above \(50\%\)), directly reflecting the significant Mg-Mg bonds/the Mg-dominated conduction bands (Figs. 6(a)). The critical orbital hybridizations might be combined with the ferromagnetic configurations, being characterized by the spin-split DOS, in which the spin-up and spin-down determine the net magnetic moment. Most of deep valence states cancel with each other, furthermore, the main contributions arise from the unbalanced electronic energy spectrum across \(E_0\), e.g., those for the \(16.7\%\) and \(3.1\%\) cases. The results show that the Mg-3s and Si-3p\(_z\) orbitals dominate the spin-split distribution, especially for the former. 

The spatial charge distributions and their variations after the Mg-chemisorptions, as clearly shown in Figs. 6(e)-6(h). The \(\pi\)-electronic bondings (Figs. 6(e) and 6(f) on the \textbf{xz}-plane) are strongly modified under Mg absorption, while the opposite is true for the \(\sigma\)-electronic ones. There exist the significant and unusual charge density differences between the guest and host atoms. The Mg-induced charge density variation is more obvious  compared with the Na-adsorption case. The main reason lies in two 3s-valence electrons for each Mg atom. The strong orbital hybridizations of 3s-3p\(_z\) in Mg-Si bonds have led to the strong modified Dirac-cone structures. Very interesting, the observable charge density variations, with positive and negative \(\Delta\rho \), are revealed between two Mg atoms under the high Mg-concentration cases (\(50\%\)), e.g., that within the black rectangle on the \textbf{yz}-plane. The strong 3s-3s orbital hybridizations in Mg-Mg bonds are responsible for the wide Mg-dominated valence and conduction energy spectra, being consistent with the atom- and orbital-decomposed DOS.

\subsection{Al adatoms and creation of diversified phenomena}

The optimal aluminum adatom adsorptions on silicene surface, as clearly shown in Fig. S3. The highest adatom adsorption can achieve through the \(50\%\) case of the single-side adsorption with the greatly enhanced buckling. Different from the Na- and Mg-adsorption cases, the \(100\%\) saturated adsorption with the double-side configuration is not stable and thus thoroughly disappears, mainly owing to the quite strong repulsive interactions among the neighboring Al adatoms. The Al-Si bond lengths, which depend on the adsorption condition, lie in the range of ${\sim\,2.58}$ \AA \ - ${2.89}$ \AA \ (Table 3). They become longer in the increase of Al-concentration. On the other side, the optimal position of aluminum-adsorbed graphene systems corresponds to the hollow site on a planar honeycomb lattice. The concentration- and configuration-dependent Al-C bond lengths are ${\sim\,2.54}$ \AA \ - ${2.57}$ \AA \ (details in \cite{41}), and the highest Al-adatom concentration, with a stable geometry, is examined to be the \(25\%\) case under the double-side adsorption.  Apparently, the geometric symmetry is responsible for the critical chemical bondings and thus the electronic and magnetic properties. 

The valley sites and the multi-orbital hybridizations in aluminum-adsorbed silicene systems can create the rich and unique band structures as shown in Fig. 7, being in sharp contrast with the Na- and Mg-adsorption cases (Figs. 2 and Fig. 5). The low-lying band structures are very complicated even at low Al-concentrations. It is very difficult to accurately identify the \(\pi\) and \(\pi^*\) bands/the \(\sigma\) bands at the sufficiently high concentrations (Figs. 7(a)-7(d)), since there exist certain conduction subbands across the Fermi level, the non-monotonous/oscillatory/partially flat energy dispersions, the Si- and/or Al-dominated valence and conduction bands. However, the separated Dirac-cone structures at the \(K\) point will be gradually recovered during the decrease of Al-concentration as marked by the red cirles. Very interesting, the Al adatoms play a critical role in the low-energy states across the Fermi level, especially for the high-concentration cases \( >25 \%\) in Figs. 7(a)-7(c). Also, they make significant contributions to the deeper valence states in the range of -5.20 eV\({\rm  <  E  < }\)-3.0 eV, in which their width is sensitive to the Al-concentration. That is to say, more energy subbands are created after aluminum adsorptions because each adatom contributes three-orbital electrons. Under certain Al-adsorption cases, the spin-split energy bands come to exist at the low energy. These electronic energy spectra clearly illustrate the ferromagnetic configuration with an observable net magnetic moment (Table 3). That is to say, the spin-configuration-induced many-particle Coulomb interactions,\cite{42} and the multi-/single-orbital hybridizations,\cite{41,42} compete/cooperate with each other to achieve the lowest ground state energy. 

The significant orbital hybridizations/the \(\pi\) and \(\sigma\) bondings, which could survive in Al-adsorbed silicene systems are thoroughly investigated from the atom-, orbital- and spin-projected van Hove singularities (Figs. 8(a)-8(f)). All the absorption cases present a finite DOS at the Fermi level, directly indicating the metallic or semimetallic behaviors. The seriously distorted V-shape structure, corresponding to the strongly modified Dirac-cone structures, is dominated by the Si-3p\(_z\) orbital (the red curve) and comes to exist below \(E_F\) only under the low Al-concentration, e.g., those for the \(12.5\%\) and \(3.1\%\) cases (Figs. 8(d) and 8(f)). This indicates the existence of the addressed \(\pi\) bonding and the n-type doping. In general, the special structures due to the four orbital of Si-host atoms could appear simultaneously within certain energy ranges of valence and conduction spectra. Such result suggests the significant/observable sp\(^3\) bonding in the Si-Si bonds. Most important, the Si-(3s, 3p\(_x\), 3p\(_y\), 3p\(_z\)) and Al-(3s, 3p\(_x\)+3p\(_y\)) orbitals (the purple, green, dashed blue, red, light blue and yellow curves) frequently present the similar van Hove singularities at the identical energies within the whole range except for a very low Al-concentration (e.g., the \(3.1\%\) case in Fig. 8(f)). Apparently, their multi-orbital hybridizations play critical roles in the diversified electronic properties. Specifically, under very high Al-concentrations, such as the \(50\%\) and 33.3\(\%\) configurations, the effective distribution widths of Al-3s/Al-[3p\(_x\)+3p\(_y\)] are more than 5 eV, clearly reflecting the creation of the Al-dominated valence and conduction bands (Figs. 8(a) and 8(b)). 

The critical factor is the strongest chemisorption due to three valence electrons in each aluminum atom. The sufficiently high carrier density can reach the intermediate region between Al and Si atoms, apparently resulting in the obvious charge redistributions of Al-(3s, 3p\(_x\)+3p\(_y\)) and Si-(3s, 3p\(_x\),3p\(_y\), 3p\(_z\)) orbitals. The prominent Al-Si bonding is also revealed as the drastic changes of carrier densities, where the large charge transfers, with the positive and negative values, simultaneously appear through the heavy red and blue regions in Fig. 9, respectively. Furthermore, the prominent Al-adsorption effects will greatly strengthen the sp\(^3\) bondings in Si-Si bonds or destroy the well characterizations of the \(\pi\) and \(\sigma\) ones. Specifically, the enough important Al-Al bonds come to exist only under the high Al-concentrations, e.g., the \(50\%\) configuration with the apparent charge variation between two adatoms (the black rectangle). On the other side, both Na-Si and Mg-Si atomic interactions are closely related to the host atoms on the different dimers (Figs. 4(a)-4(c) and Figs. 6(e) and 6(f)), where the carrier density is very dilute at the intermediate region close to the Na/Mg adatoms. The important difference with the Al-adsorption cases arise from the hollow-site absorption position and one-/two-3s valence electrons in Na/Mg.
 
\subsection{Available experimental measurement and potential applications}

ARPES measurements are very successfully in identifying the diverse band structures of graphene systems, e.g., the gapless valence Dirac cone in monolayer graphene,\cite{43} two pairs of parabolic energy dispersions in AB-stacked bilayer system,\cite{43,44} the coexistent linear and parabolic bands in twisted bilayer graphenes.\cite{45} Moreover, the Na-/Li-adatom adsorptions\cite{47,48} on graphene surfaces have been verified to present the blue shift of the Fermi level and the linear Dirac cone. In case of silicene, the linear dispersion with the Fermi velocity of v$_F$=1.3x10$^6$ ms$_1$ comparable to graphene's has been observed.\cite{32,46} The similar ARPES examinations could be done for (Na,Mg,Al)-adsorbed silicenes to explore the blue shift of the Fermi level, the low-lying energy spectra with the modified Dirac cones and the adatom-dominated bands, and the splitting middle-energy valence bands near the M point. As a result, they are useful in estimating the partial charge transfer from guest to host atoms and understanding the modified \(\pi\) bonding.  In addition, the optical\cite{49} and transport\cite{50} measurements might be efficient and reliable in examining the (Na, Mg, Al)-created free electron density. 

The (Na,Mg,Al)-adsorption-enriched van Hove singularities could be directly verified from the high-resolution STS measurements, as successfully done for carbon-based materials with the well-characterized sp\(^2\) bondings. For example, the up-to-date experiments on layered graphene systems have clearly identified an isotropic V-shape structure vanishing at the gapless Dirac point for monolayer graphene,\cite{43} the zone-folding-created peaks in the logarithmic form for twisted bilayer graphenes,\cite{45} a delta-function-like peak centered about E$_F$ for ABC stackings,\cite{51} the semi-metallic property for graphite,\cite{44,48} and its splitting \(\pi\) and \(\pi^*\) strong peaks at deeper/higher state energies.\cite{44} Morever, STS has sucessfully confirmed the existence of Dirac fermions in silicene.\cite{51} The further STS examinations for (Na,Mg,Al)-adsorbed silicene systems are mainly focused on the modified dip structure below the Fermi level, the split \(\pi\) and \(\pi^*\) peaks, and the adatom-generated extended structure. On the other hand, the theoretical predictions on the (Mg,Al)-adatom-generated ferromagnetism could be directly test by using the high-resolution spin-polarized-STM.

Recently, silicene-related materials have attracted great attention as promising materials for Li/Na-ions batteries with high theoretical capacity through electrochemical conversion reactions and low-cost production.\cite{11,12,52,53} Further approaches to enhance the application of silicene as anode materials in LIBs have been carried out, such as compositing by doping. It is reported that with Na modification, performance of silicene could achieve high theoretical capacity of 954 mAh/g and a low diffusion barrier of 0.23 eV.\cite{11} As compared to graphite, Na interacts with silicene leads to a higher electrode potential, which suppresses dendritic Na growth.\cite{11,55} In addition, owning to the strong spin-orbit coupling and easily tunable bandgap at Dirac-cone, silicene-based materials can successfully be used in spintronic devives,\cite{13,56,57} and FET at room temperature.\cite{9,10} Silicene materials are expected to superior graphene on real applications based on the existing silicon-based industry. Further systematic studies are needed to find out the most suitable absorbers on silicene as high potential in the nanoscale electronic devices.

\section{IV. Conclusions and Summary}

 Different from the tiny-gap semiconductor behavior of pristine silicene, all the metal-adsorbed cases belong to the n-type dopings, implying the enhanced electronic conductivity. On the other hand, Na-, Mg- and Al-adsorbed silicenes quite differ from one another in the essential properties. The important differences cover the optimal adsorption sites, distinct adatom-silicon bond lengths, the enhanced buckling degrees, the free carrier densities, the modified Dirac-cone structures, the well-behaved or undefined \(\pi\) and \(\sigma\) valence states (the former and the latter in the Na- and Mg-/Al-adsorbed silicenes), the spin-degenerate or spin-split electronic energy spectra across \(E_F\), the adatom-enriched charge density distributions, the ferromagnetic or non-magnetic spin distributions under certain adsorption configurations, and the adsorption-diversified van Hove singularities. The critical mechanisms are deduced to arise from the distinct single-/multi-orbital hybridizations in adatom-Si, Si-Si and adatom-adatom bonds. Most important, the Na-Si, Mg-Si and Al-Si bonds, which account for the diverse physical phenomena, are deduced to possess the 3s-3p$_z$, 3s-3p$_z$ and (3s,3p$_x$+3p$_y$)-(3s,3p$_x$,3p$_y$,3p$_z$) orbital hybridizations, respectively. The Al-adsorption configurations present the most significant chemical modifications, such as more oscillatory and partially flat energy dispersions across the Fermi level, the thorough disappearance of the modified Dirac-cone structures at the sufficiently high adatom concentrations, and the largest charge density difference in the Al-Si and Al-Al bonds. The ferromagnetic configurations are shown to only survive in the specific Mg- and Al-adsorptions, but not the Na-cases. This indicates that the spin-dependent many-particle interactions might play a critical role in reducing the total ground state energy. The guest- or host-atom-dominated spin distributions are consistent with the spin-split DOS across the Fermi level. The present work should serve as a first step towards further investigation into other necessary properties of metal-adsorbed silicnene for fabrication and potential device applications. The similar metal-adatom adsorption effects could be generalized to Ge-, Sn- and Pb-monolayers.
 
 \par\noindent {\bf Conflict of interest}
   
   There are no conflicts of interest in this paper

  \par\noindent {\bf Acknowledgments}
     
 This work is supported by the Hi-GEM Research Center and the Taiwan Ministry of Science and Technology under grant number MOST 108-2112-M-006-022-MY3.

 %\bibliography{achemso}

\begin{thebibliography}{99}
  
\bibitem{1} Borenstein, A., Hanna, O., Attias, R., Luski, S., Brousse, T., \& Aurbach, D. Carbon-based composite materials for supercapacitor electrodes: a review.  \textit{Journal of Materials Chemistry A}, \textbf{5}(25), 12653-12672(2017). 

\bibitem{2}  Yu, X., Cheng, H., Zhang, M., Zhao, Y., Qu, L., \& Shi, G. Graphene-based smart materials. \textit{Nature Reviews Materials}, \textbf{2}(9), 1-13(2017).

\bibitem{3} Oughaddou, H. \textit{et al}. Silicene, a promising new 2D material. \textit{Progress in Surface Science}, \textbf{90}(1), 46-83(2015).

\bibitem{4} Shao, Z. G., Ye, X. S., Yang, L., \& Wang, C. L. First-principles calculation of intrinsic carrier mobility of silicene. \textit{Journal of Applied Physics}, \textbf{114}(9), 093712(2013). 

\bibitem{5} Mokkath, J. H., \& Schwingenschlögl, U. Tunable optical absorption in silicene molecules. \textit{Journal of Materials Chemistry C}, \textbf{4}(31), 7387-7390(2016).

\bibitem{6} Liu, C. C., Feng, W., \& Yao, Y. Quantum spin Hall effect in silicene and two-dimensional germanium. \textit{Physical review letters}, \textbf{107}(7), 076802(2011).

\bibitem{7} Drummond, N. D., Zolyomi, V., \& FalKo, V. I. Electrically tunable band gap in silicene. \textit{Physical Review B}, \textbf{85}(7), 075423(2012).   

\bibitem{8}  Aftab, T. Valleytronics and phase transition in silicene. \textit{Physics Letters A}, \textbf{381}(10), 935-943(2017).

\bibitem{9} Vali, M., Dideban, D., \& Moezi, N. Silicene field effect transistor with high on/off current ratio and good current saturation. \textit{Journal of Computational Electronics}, textbf{15}(1), 138-143(2016).

\bibitem{10}Tao, L. \textit{et al}. Silicene field-effect transistors operating at room temperature. \textit{Nature nanotechnology}, \textbf{10}(3), 227-231(2015).

\bibitem{11}  Zhu, J., \& Schwingenschlogl, U. Silicene for Na-ion battery applications. \textit{2D Materials}, \textbf{3}(3), 035012(2016).

\bibitem{12}  Xu, S., Fan, X., Liu, J., Singh, D. J., Jiang, Q., \& Zheng, W.  Adsorption of Li on single-layer silicene for anodes of Li-ion batteries. \textit{Physical Chemistry Chemical Physics}, \textbf{20}(13), 8887-8896(2018).

\bibitem{13} Ezawa, M. Spin valleytronics in silicene: Quantum spin Hall–quantum anomalous Hall insulators and single-valley semimetals. \textit{Physical Review B}, \textbf{87}(15), 155415(2013).

\bibitem{14}  Osborne, D. A., Morishita, T., Tawfik, S. A., Yayama, T., \& Spencer, M. J. Adsorption of toxic gases on silicene/Ag (111). \textit{Physical Chemistry Chemical Physics}, \textbf{21}(32), 17521-17537(2019).

\bibitem{15} Vogt, P. \textit{et al}. Silicene: compelling experimental evidence for graphenelike two-dimensional silicon. \textit{Physical review letters}, \textbf{108}(15), 155501(2012).

\bibitem{16} Meng, L. \textit{et al} Buckled silicene formation on Ir (111). \textit{Nano letters}, \textbf{13}(2), 685-690(2013).

\bibitem{17} Aizawa, T., Suehara, S., \& Otani, S. (2015). Phonon dispersion of silicene on ZrB2 (0 0 0 1). \textit{Journal of Physics: Condensed Matter}, \textbf{27}(30), 305002.

\bibitem{18} J. Zhu and U. Schwingenschlögl, Silicene on MoS2: role of the van der Waals interaction. \textit{2D Mater.}, \textbf{2}, 045004(2015).

\bibitem{19} H. Liu, J. Gao and J. Zhao, Silicene on Substrates: A Way to Preserve or Tune its Electronic Properties. \textit{J. Phys. Chem. C}, \textbf{117}, 10353–10359(2013).

\bibitem{31}  Pan, Y. \textit{et al} Construction of 2D atomic crystals on transition metal surfaces: graphene, silicene, and hafnene. \textit{Small}, \textbf{10}(11), 2215-2225(2014).

\bibitem{20}  Jose, D and Datta, A. Structures and chemical properties of silicene: unlike graphene.  \textit{Accounts of Chemical Research},  \textbf{47}, 593-602(2013).

\bibitem{21} Lin, S. Y., Chang, S. L., Tran, N. T. T., Yang, P. H and Lin, M. F.  H-Si bonding-induced unusual electronic properties of silicene: a method to identify hydrogen concentration. \textit{Physical Chemistry Chemical Physics},   \textbf{17}, 26443-26450(2015).

\bibitem{22} Nguyen, D. K., Tran, N. T. T., Chiu, Y. H and Lin, M. F. Concentration-Diversified Magnetic and Electronic Properties of Halogen-Adsorbed Silicene. \textit{Scientific Reports}, \textbf{9}, 1-15(2019).

\bibitem{23} Du, Y \textit{et al}. Tuning the band gap in silicene by oxidation. \textit{ACS Nano}, \textbf{8}, 10019-10025(2014).

\bibitem{24} Xu, K., Ben, L., Li, H., \& Huang, X.  Silicon-based nanosheets synthesized by a topochemical reaction for use as anodes for lithium ion batteries. \textit{Nano Research}, \textbf{8}(8), 2654-2662(2015).

\bibitem{25} Wu, C. H. Integer quantum Hall conductivity and longitudinal conductivity in silicene under the electric field and magnetic field. \textit{The European Physical Journal B}, \textbf{92}(2), 25(2019).

\bibitem{26} Z. Ni, Q. Liu, K. Tang, J. Zheng, J. Zhou, R. Qin, Z. Gao, D. Yu and J. Lu, Tunable bandgap in silicene and germanene, \textit{Nano Lett.},  \textbf{12}, 113–118(2012).

\bibitem{27} Mirershadi, S., \& Sattari, F. Spin and valley dependent transport in strained silicene superlattice. \textit{Physica E: Low-dimensional Systems and Nanostructures},  \textbf{115}, 113696(2020).

\bibitem{28} Xie, H., Ouyang, T., Germaneau, É., Qin, G., Hu, M., \& Bao, H. Large tunability of lattice thermal conductivity of monolayer silicene via mechanical strain.  \textit{Physical Review B}, \textbf{93}(7), 075404(2016).

\bibitem{29} Sahin, H., \& Peeters, F. M. Adsorption of alkali, alkaline-earth, and 3 d transition metal atoms on silicene. \textit{Physical Review B}, \textbf{87}(8), 085423(2013). .

\bibitem{30} Xu, Q., Yang, G. M., Fan, X., \& Zheng, W. T. Adsorption of metal atoms on silicene: stability and quantum capacitance of silicene-based electrode materials. \textit{Physical Chemistry Chemical Physics}, \textbf{21}(8), 4276-4285(2019).    

\bibitem{32}  Avila, J. \textit{et al}. Presence of gapped silicene-derived band in the prototypical (3x3) silicene phase on silver (111) surfaces. \textit{Journal of Physics: Condensed Matter}, \textbf{25}(26), 262001(2013).

\bibitem{33} Siegel, D. A., Regan, W., Fedorov, A. V., Zettl, A and Lanzara, A.  Charge-carrier screening in single-layer graphene. \textit{Physical Review Letters},   \textbf{110}, 146802(2013).

\bibitem{34} Kresse, G and  Furthm{\"u}ller, J.  Efficient iterative schemes for ab initio total-energy calculations using a plane-wave basis set. \textit{Physical Review B},   \textbf{54}, 11169(1996).

\bibitem{35} Kresse, G and Joubert, D. From ultrasoft pseudopotentials to the projector augmented-wave method. \textit{Physical Review B},  \textbf{59}, 1758(1999).

\bibitem{36}  Perdew, J. P., Burke, K and  Ernzerhof, M.   Generalized gradient approximation made simple. \textit{Physical Review Letters}, \textbf{77},  3865(1996).

\bibitem{37}  Bl{\"o}chl, P. E. Projector augmented-wave method. \textit{Physical Review B},   \textbf{50},17953(1994).

\bibitem{38}  Grimme, S.  Semiempirical GGA-type density functional constructed with a long-range dispersion correction. \textit{Journal of Computational Chemistry},   \textbf{27},  1787-1799(2006).

\bibitem{38a} Ni, Z. \textit{et al}. Tunable band gap and doping type in silicene by surface adsorption: towards tunneling transistors. \textit{Nanoscale}, \textbf{6}(13), 7609-7618(2014).

\bibitem{38b} Botari, T., Perim, E., Autreto, P. A. S., Van Duin, A. C. T., Paupitz, R., \& Galvao, D. S. Mechanical properties and fracture dynamics of silicene membranes. \textit{Physical Chemistry Chemical Physics}, \textbf{16}(36), 19417-19423(2014).

\bibitem{39}  Zhang, X. L., Liu, L. F and Liu, W. M.  Quantum anomalous Hall effect and tunable topological states in 3d transition metals doped silicene.  \textit{Scientific Reports},   \textbf{3},  2908(2013). 

\bibitem{40}  Jin, H., Jeong, H., Ozaki, T and Yu, J. Anisotropic exchange interactions of spin-orbit-integrated states in Sr\(_2\)IrO\(_4\). \textit{Physical Review B},   \textbf{80},  075112(2009).

\bibitem{41} Lin, S. Y., Lin, Y. T., Tran, N. T. T., Su, W. P and Lin,  M. F. Feature-rich electronic properties of aluminum-absorbed graphenes. \textit{Carbon},   \textbf{120}, 209-218(2017).

\bibitem{42} Wu, Y. P., Rahm, E and Holze, R. Carbon anode materials for lithium ion batteries. \textit{Journal of Power Sources},  \textbf{114}, 228-236(2003).

\bibitem{43} Ohta, T., Bostwick, A., McChesney, J. L., Seyller, T., Horn, K and Rotenberg, E.  Interlayer interaction and electronic screening in multilayer graphene investigated with angle-resolved photoemission spectroscopy. \textit{Physical Review Letters},   \textbf{98},  206802(2007).   

\bibitem{44}  Ohta, T., Bostwick, A., Seyller, T., Horn, K and Rotenberg, E.   Controlling the electronic structure of bilayer graphene.  \textit{Science},   \textbf{313},  951954(2006).

\bibitem{45}   Zhou, S. Y., Siegel, D. A., Fedorov, A. V and Lanzara, A.  Metal to insulator transition in epitaxial graphene induced by molecular doping. \textit{Physical Review Letters},   \textbf{101}, 086402(2008).     

\bibitem{47}  Papagno, M \textit{et al}. Large band gap opening between graphene Dirac cones induced by Na adsorption onto an Ir superlattice. \textit{ACS Nano}, \textbf{6},  199204(2011).

\bibitem{48}   Sugawara, K., Kanetani, K., Sato, T and Takahashi, T.   Fabrication of Li-intercalated bilayer graphene.  \textit{AIP Advance},   \textbf{1}, 022103(2011).        

\bibitem{46}Vogt, P.\textit{et al}. Silicene: compelling experimental evidence for graphenelike two-dimensional silicon. \textit{Physical review letters}, \textbf{108}(15), 155501(2012).

\bibitem{49}   Borensztein, Y., Pr{\'e}vot, G and Masson, L.  Large differences in the optical properties of a single layer of Si on Ag (110) compared to silicene. \textit{Physical Review B},   \textbf{89}, 245410(2014).  

\bibitem{50}  Ao, L \textit{et al}. Silicene field-effect transistors operating at room temperature.  \textit{Nature Nanotechnology}, \textbf{10}, 227(2015).

\bibitem{51} Chen, L. \textit{et al}. Evidence for Dirac fermions in a honeycomb lattice based on silicon. \textit{Physical review letters}, \textbf{109}(5), 056804(2012).

\bibitem{52}   Zhuang, J., Xu, X., Peleckis, G., Hao, W., Dou, S. X and  Du, Y. Silicene: a promising anode for lithium-ion batteries. \textit{Advanced Materials},  \textbf{29},  1606716(2017).

\bibitem{53} Tritsaris, G. A., Kaxiras, E., Meng, S and Wang, E. Adsorption and diffusion of lithium on layered silicon for Li-ion storage. \textit{Nano Letters}, \textbf{13},  2258-2263(2013).

\bibitem{55} Persson, K., Hinuma, Y., Meng, Y. S., Van der Ven, A and Ceder, G.  Thermodynamic and kinetic properties of the Li-graphite system from first-principles calculations.  \textit{Physical Review B},   \textbf{82}, 125416(2010).

\bibitem{56}   Wang, Y  \textit{et al}. Half-metallic silicene and germanene nanoribbons: towards high-performance spintronics device.  \textit{Nano},   \textbf{7}, 1250037(2012).

\bibitem{57} Yang-Yang, W., Ru-Ge, Q., Da-Peng, Y and Jing, L. Silicene spintronics-A concise review. \textit{Chinese Physics B},  \textbf{24},  087201(2015).
            
\end{thebibliography}

\newpage
\begin{table}[htb]
	\caption{The calculated nearest passivated Si-Si and Na-Si bond lengths, Na heights, total magnetic moments, buckling and binding energies of Na-absorbed silicene systems.}
	\label{table1}
	\begin{center}                       
		\begin{tabular}{llllllll}
			\hline
			& \makecell{Na:Si} & \makecell{Passivated \\ Si-Si bond \\ length (\AA)} & \makecell{Na-Si bond \\ length (\AA)}  & \makecell{Na height\\(\AA)} & \makecell{Magnetic \\moment\\(\(\mu _B \))}&  \makecell{Buckling\\(\AA)}& \makecell{\(E_b\)\\(eV)} \\ 
			\hline
			& Pristine & 2.28 & X & X & 0 & 0.48 & X \\ 
			\hline
			& 6:6=100\(\%\) & 2.41 & 2.94 & 2.08 & 0 & 0.86 & -1.74 \\ 
			\hline                
			& 3:6=50\(\%\) & 2.33 & 3.15 & 2.28 & 0 & 0.62 & -3.03\\                                                                                    
			\hline                
			& 2:6=33.3\(\%\) & 2.31 & 3.10 & 2.18 & 0 & 0.58 & -1.48\\                  
			\hline                
			& 2:8=25\(\%\) & 2.30 & 3.10 & 2.12 & 0 & 0.57 & -2.14\\                               
			\hline                
			& 1:6=16.7\(\%\) & 2.31 & 3.12 & 2.17 & 0 & 0.52 & -1.44\\                                                                                              
			\hline                
			& 1:8=12.5\(\%\) & 2.30 & 3.17 & 2.22 & 0 & 0.48 & -1.60\\                                          
			\hline                
			& 1:18=5.6\(\%\) &2.30  &3.19  &2.26  &  0& 0.48& -1.68\\ 
			\hline                
			& 1:32=3.1\(\%\) & 2.30 & 3.26 & 2.35 & 0 & 0.48 & -1.86\\ 
			\hline                                                                                
		\end{tabular}
	\end{center}
\end{table} 
\newpage

\begin{table}[htb]
	\caption{The calculated nearest passivated Si-Si and Mg-Si bond lengths, Mg heights, total magnetic moments, buckling and binding energies of Mg-absorbed silicene systems.}
	\label{table2}
	\begin{center}                       
		\begin{tabular}{llllllll}
			\hline
			& \makecell{Mg:Si} & \makecell{Passivated \\ Si-Si bond \\ length (\AA)} & \makecell{Mg-Si bond \\ length (\AA)}  & \makecell{Mg height\\(\AA)} & \makecell{Magnetic \\moment\\(\(\mu _B \))}&  \makecell{Buckling\\(\AA)}& \makecell{\(E_b\)\\(eV)} \\ 
			\hline
			& 6:6=100\(\%\) & 2.46 & 2.86 & 2.24 & 0 & 0.94 & -0.77 \\ 
			\hline                
			& 3:6=50\(\%\) & 2.43 & 2.81 & 2.08 & 0 & 0.86 & -1.66\\                                                                                    
			\hline                
			& 2:6=33.3\(\%\) & 2.38 & 2.81 & 2.09 & 0 & 0.79 & -1.33\\                  
			\hline                
			& 2:8=25\(\%\) & 2.40 & 2.70 & 1.95 & 0 & 0.73 & -1.67\\                               
			\hline                
			& 1:6=16.7\(\%\) & 2.33 & 2.76 & 1.89 & 0.46 & 0.61 & -0.92\\                                                                                              
			\hline                
			& 1:8=12.5\(\%\) & 2.33 & 2.81 & 1.94 & 0 & 0.59 & -0.98\\                                          
			\hline                
			& 1:18=5.6\(\%\) & 2.31 &  2.87& 2.08 & 0 &0.53 & -0.92\\ 
			\hline                
			& 1:32=3.1\(\%\) & 2.31 & 2.88 & 2.12 & 0.53 & 0.52 & -0.82\\ 
			\hline                                                                                
		\end{tabular}
	\end{center}
\end{table}                                             
\vskip0.5 truecm

\begin{table}[htb]
	\caption{The calculated nearest passivated Si-Si and Al-Si bond lengths, Al- heights, total magnetic moments, and binding energies of Al-absorbed silicene systems.}
	\label{table3}
	\begin{center}                       
		\begin{tabular}{llllllll}
			\hline
			& \makecell{Al:Si} & \makecell{Passivated \\ Si-Si bond \\ length (\AA))} & \makecell{Al-Si bond \\ length (\AA)}  & \makecell{Al height\\(\AA)} & \makecell{Magnetic \\moment\\(\(\mu _B \))}&  \makecell{Buckling\\(\AA)}& \makecell{\(E_b\)\\(eV)} \\ 
			\hline                
			& 3:6=50\(\%\) & 2.47 & 2.89 & 2.14 & 0 & 0.74 & -3.11\\                                                                                    
			\hline                
			& 2:6=33.3\(\%\) & 2.47 & 2.71 & 1.93 & 0 & 0.70 & -2.69\\                  
			\hline                
			& 2:8=25\(\%\) & 2.47 & 2.72 & 2.05 & 0.34 & 0.75 & -2.73\\                               
			\hline                
			& 1:6=16.7\(\%\) & 2.39 & 2.61 & 1.90 & 0 & 0.75 & -3.10\\                                                                                              
			\hline                
			& 1:8=12.5\(\%\) & 2.34 & 2.60 & 1.89 & 0.73 & 0.64 & -3.47\\                                          
			\hline                
			& 1:18=5.6\(\%\) & 2.33 &2.57  & 1.86 & 0 &0.57 &-3.54 \\ 
			\hline                
			& 1:32=3.1\(\%\) & 2.32 & 2.58 & 1.84 & 0 & 0.49 & -3.72\\ 
			\hline                                                                                
		\end{tabular}
	\end{center}
\end{table}

  \centerline {\Large \textbf {Figure Captions}}
  
\vskip0.5 truecm
        
      Figure 1: The optimal geometries of the sodium-adsorbed silicenes with the side and top views for the distinct adsorptions: (a) [3:6], (b) [2:6], (c) [1:6], (d) [1:8], (e) [1:18], and (f) [1:32].

     \vskip0.5 truecm
     
      Figure 2: The concentration- and configuration-dependent band structures of sodium-adsorbed silicene materials with (a) Pristine silicene, (b) Na:Si=[6:6], (c) [3:6], (d) [2:6], (e) [1:6], (f) [1:8], (g) [1:18] and (h) [1:32]. The Na and passivated Si dominances are illustrated as the pink and grey circles.

      \vskip0.5 truecm
      
     Figure 3: The atom- and orbital-decomposed DOS for the Na-adsorbed silicene systems: (a) Pristine silicene, (b) Na:Si=[6:6], (c) [3:6], (d) [2:6], (e) [1:6], (f) [1:8], (g) [1:18], and (h) [1:32].

      \vskip0.5 truecm
      
     Figure 4: The spatial charge distribution for (a) pristine silicene, (b) [3:6], (d) [1:8] and (f) [1:18]. Their corresponding variations after Na-chemisorptions along the \textbf{xz}- and \textbf{yz}-planes are shown in (c), (e), and (g). The \(\pi\)-bonding variations are indicated by the black arrow, whereas the Na-Na and Na-Si interactions are shown by black and red rectangles, respectively.
      
      \vskip0.5 truecm
      
     Figure 5: The concentration- and configuration-dependent band structures of magnesium-adsorbed silicene materials with the Mg and Si dominances (the blue and grey circles): (a) [3:6], (b) [1:6], (c) [1:18], and (d) [1:32]. The spin-up and spin-down states in (b) and (d) are denoted as the black and green solid curves. The corresponding spin density distributions are clearly shown on the top- and side-views.
      
      \vskip0.5 truecm
      
      Figure 6: The atom-, orbital- and spin-projected van Hove singularities for the Mg-adsorbed silicene materials under the (a) [6:6], (b) [1:6], (c) [1:18] and (d) [1:32] cases. The spatial charge distributions/the variations after Mg adsorptions on the \textbf{xz}- and \textbf{yz}- planes (e)-(h).

      \vskip0.5 truecm
      
     Figure 7: The concentration- and configuration-enriched energy bands for the Al-adsorbed silicene materials, being combined with the Al and Si contributions (the purple and grey circles): (a) [3:6], (b) [2:6], (c) [1:6], (d) [1:8], (e) [1:18] and (f) [1:32]. 

\vskip0.5 truecm
      
      Figure 8: The atom-, orbital- and spin-projected DOS for the Al-adsorbed silicenes under the distinct absorption cases: (a) [3:6], (b) [2:6], (c) [1:6], (d) [1:8], (e) [1:18] and (f) [1:32]. 
      
\vskip0.5 truecm
      
      Figure 9: The spatial charge distribution/their variations after Al-chemisorptions along the \textbf{xz}- and \textbf{yz}-planes: (a)/(b) [3:6], (c)/(d) [1:8], and (e)/(f) [1:18].

\newpage   
\begin{figure}[hp]
	\graphicspath{{figure}}
	\centering
	\includegraphics[scale=0.6]{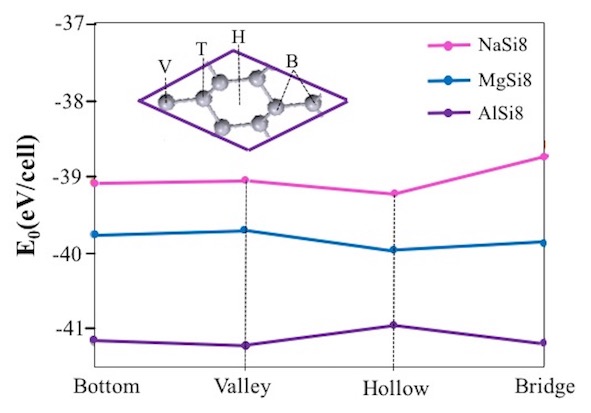}
	\caption{(Color online) Total ground state energy variations toward different adsorption position of Na/Mg/Al on silicene.}
	\label{FIG:1}
\end{figure}

\begin{figure}[hp]
	\graphicspath{{figure}}
	\centering
	\includegraphics[scale=2]{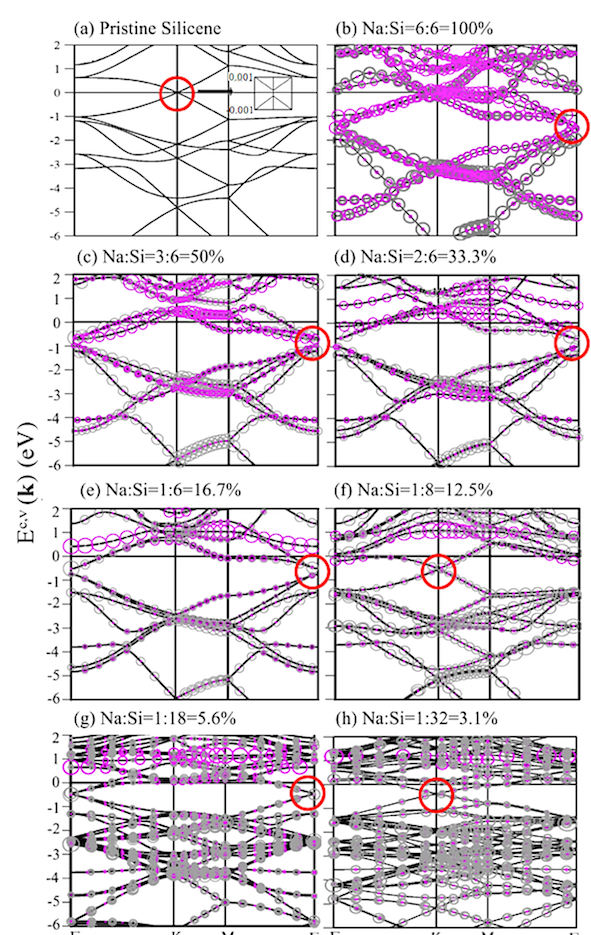}
	\caption{(Color online) The concentration- and configuration-dependent band structures of sodium-adsorbed silicene materials with (a) Pristine silicene, (b) Na:Si=[6:6], (c) [3:6], (d) [2:6], (e) [1:6], (f) [1:8], (g) [1:18] and (h) [1:32]. The Na and passivated Si dominances are illustrated as the pink and grey circles.}
	\label{FIG:2}
\end{figure}       

\begin{figure}[hp]
	\graphicspath{{figure}}
	\centering
	\includegraphics[scale=0.3]{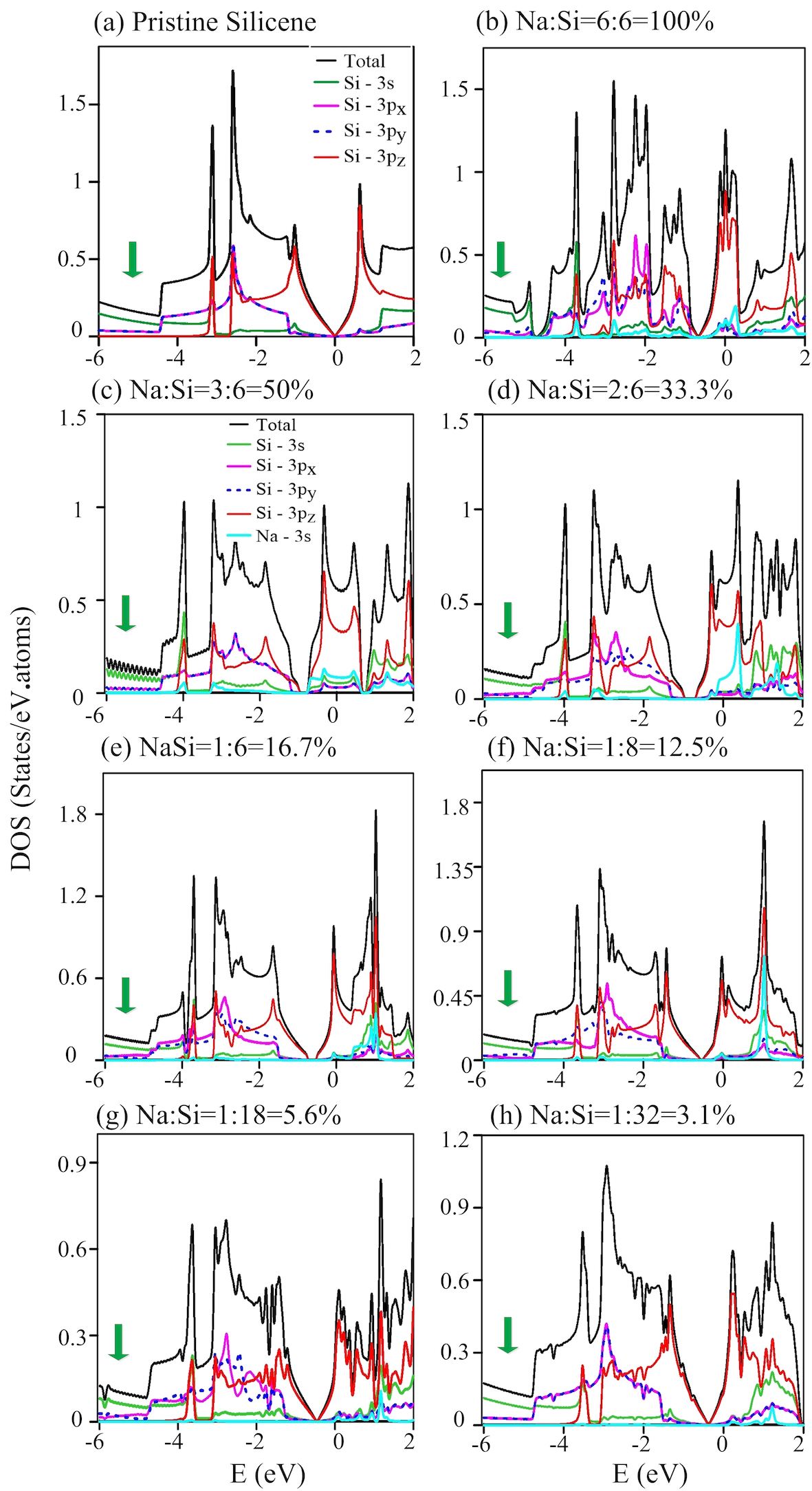}
	\caption{(Color online) The atom- and orbital-decomposed DOS for the Na-adsorbed silicene systems: (a) Pristine silicene, (b) Na:Si=[6:6], (c) [3:6], (d) [2:6], (e) [1:6], (f) [1:8], (g) [1:18], and (h) [1:32].}
	\label{FIG:3}
\end{figure}  

\begin{figure}[hp]
	\graphicspath{{figure}}
	\centering
	\includegraphics[scale=0.7]{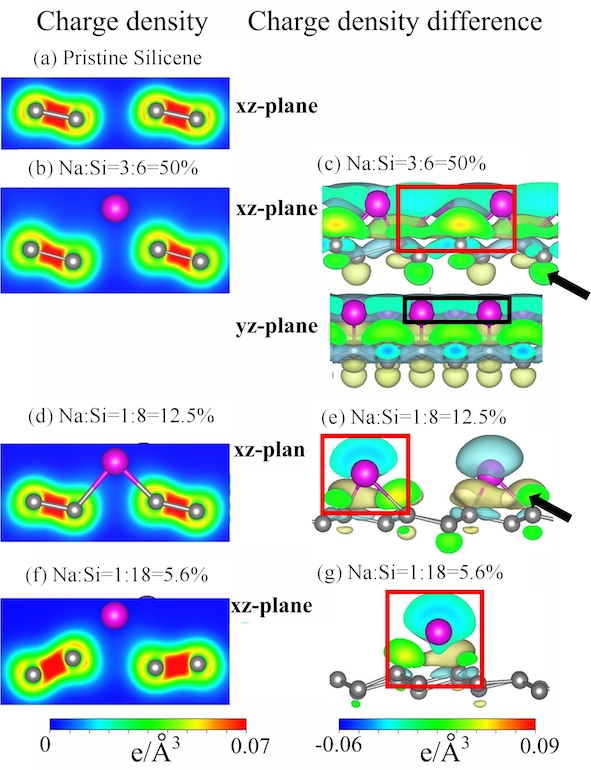}
	\caption{(Color online) The spatial charge distribution for (a) pristine silicene, (b) [3:6], (d) [1:8] and (f) [1:18]. Their corresponding variations after Na-chemisorptions along the \textbf{xz}- and \textbf{yz}-planes are shown in (c), (e), and (g). The \(\pi\)-bonding variations are indicated by the black arrow, whereas the Na-Na and Na-Si interactions are shown by black and red rectangles, respectively.}
	\label{FIG:4}
\end{figure}                             

\begin{figure}[hp]
	\graphicspath{{figure}}
	\centering
	\includegraphics[scale=0.3]{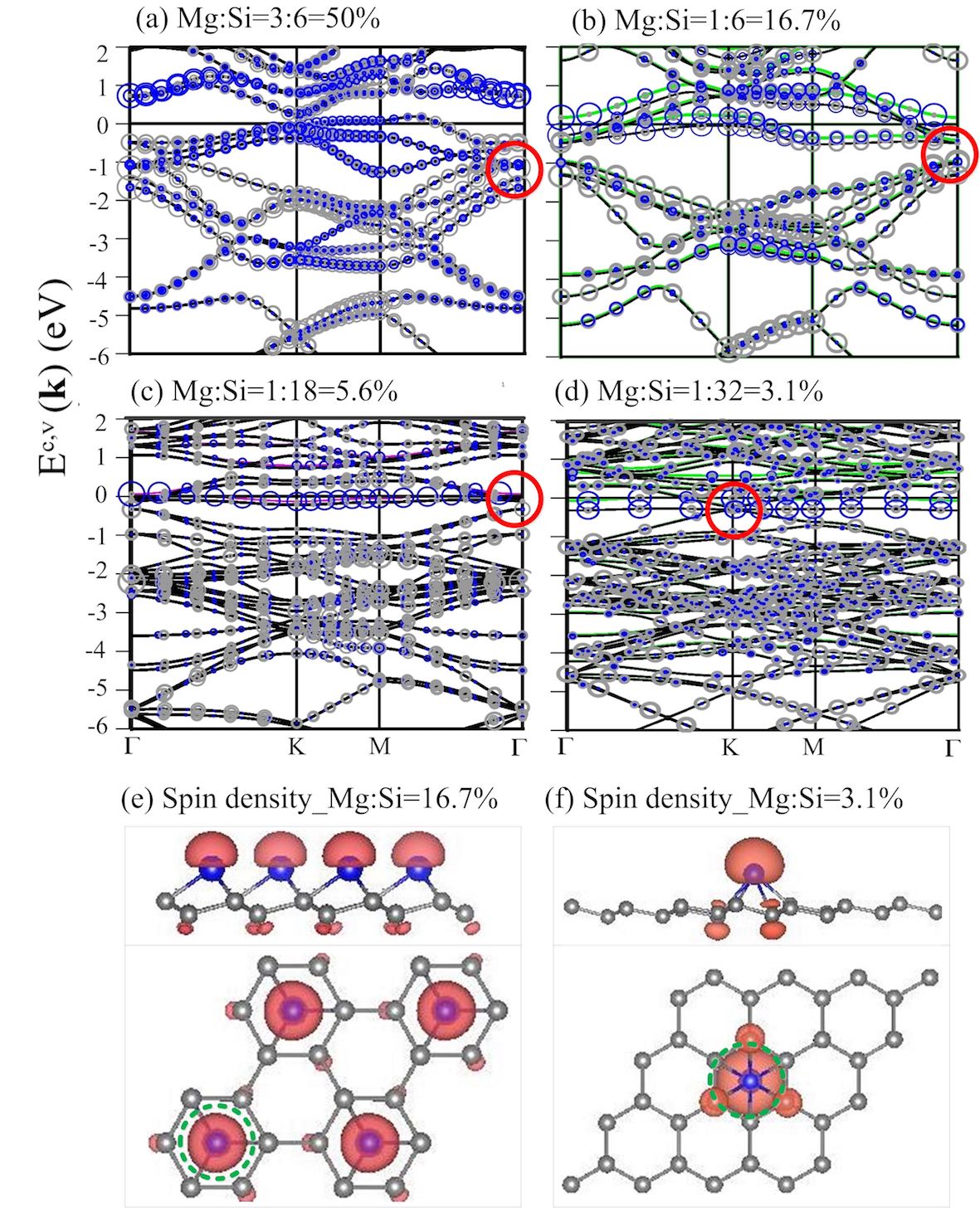}
	\caption{(Color online) The concentration- and configuration-dependent band structures of magnesium-adsorbed silicene materials with the Mg and Si dominances (the blue and grey circles): (a) [3:6], (b) [1:6], (c) [1:18], and (d) [1:32]. The spin-up and spin-down states in (b) and (d) are denoted as the black and green solid curves. The corresponding spin density distributions are clearly shown on the top- and side-views.}
	\label{FIG:5}
\end{figure}   

\begin{figure}[hp]
	\graphicspath{{figure}}
	\centering
	\includegraphics[scale=0.25]{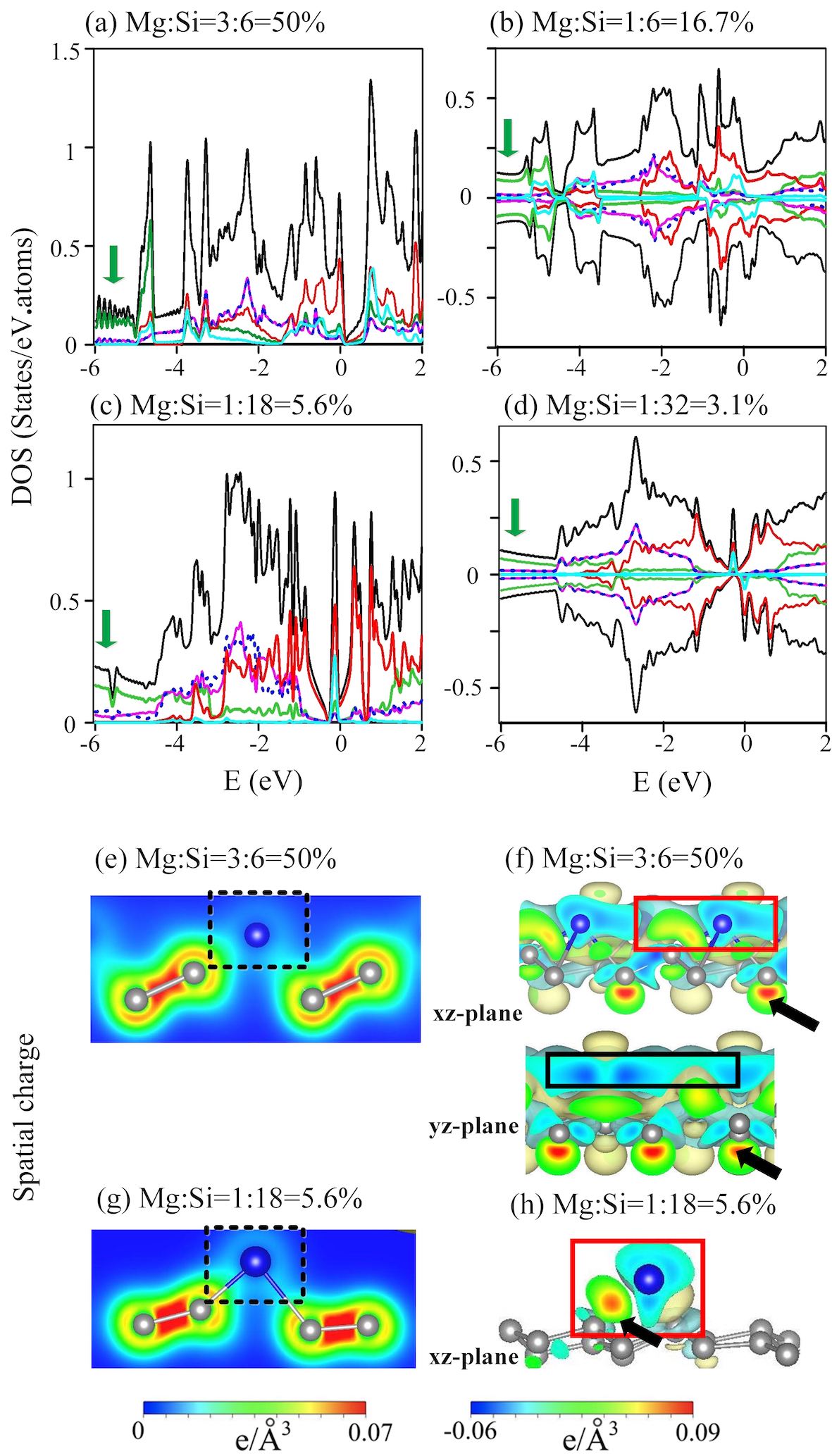}
	\caption{(Color online) The atom-, orbital- and spin-projected van Hove singularities for the Mg-adsorbed silicene materials under the (a) [6:6], (b) [1:6], (c) [1:18] and (d) [1:32] cases. The spatial charge distributions/the variations after Mg adsorptions on the \textbf{xz}- and \textbf{yz}- planes (e)-(h).}
	\label{FIG:6}
\end{figure}   

\begin{figure}[hp]
	\graphicspath{{figure}}
	\centering
	\includegraphics[scale=0.35]{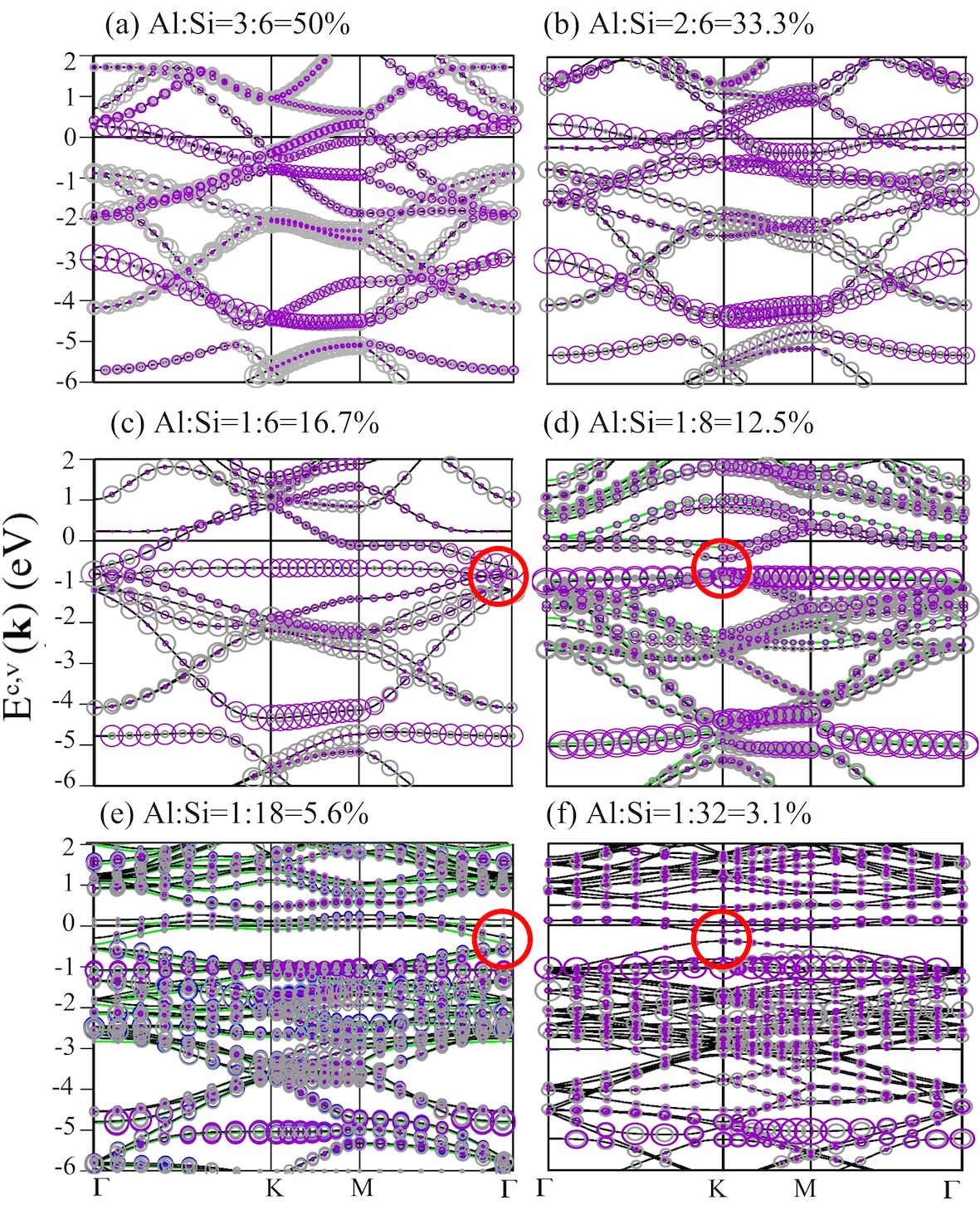}
	\caption{(Color online) The concentration- and configuration-enriched energy bands for the Al-adsorbed silicene materials, being combined with the Al and Si contributions (the purple and grey circles): (a) [3:6], (b) [2:6], (c) [1:6], (d) [1:8], (e) [1:18] and (f) [1:32]. }
	\label{FIG:7}
\end{figure}  

\begin{figure}[hp]
	\graphicspath{{figure}}
	\centering
	\includegraphics[scale=0.35]{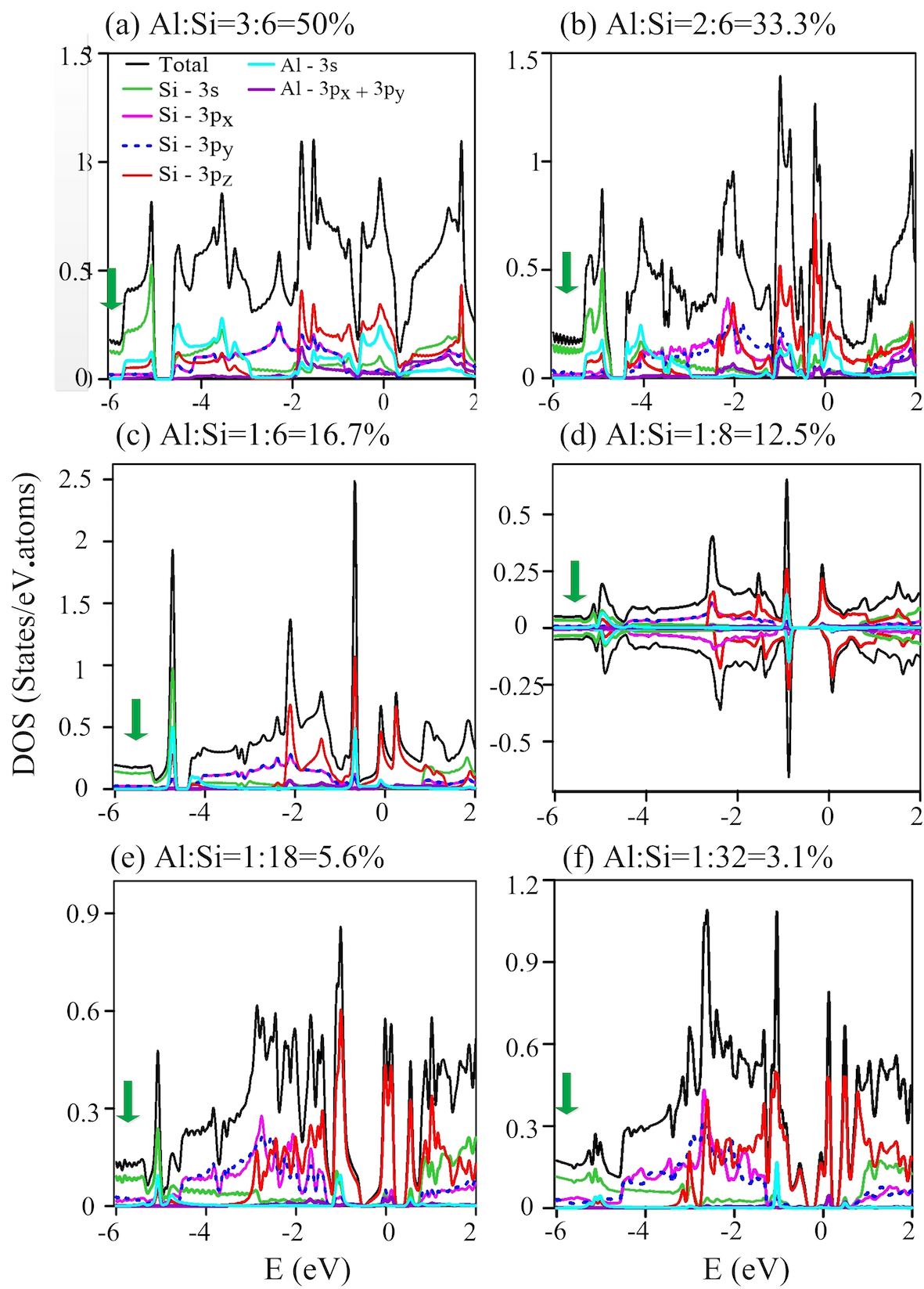}
	\caption{(Color online) The atom-, orbital- and spin-projected DOS for the Al-adsorbed silicenes under the distinct absorption cases: (a) [3:6], (b) [2:6], (c) [1:6], (d) [1:8], (e) [1:18] and (f) [1:32]. }
	\label{FIG:8}
\end{figure}      

\begin{figure}[hp]
	\graphicspath{{figure}}
	\centering
	\includegraphics[scale=0.7]{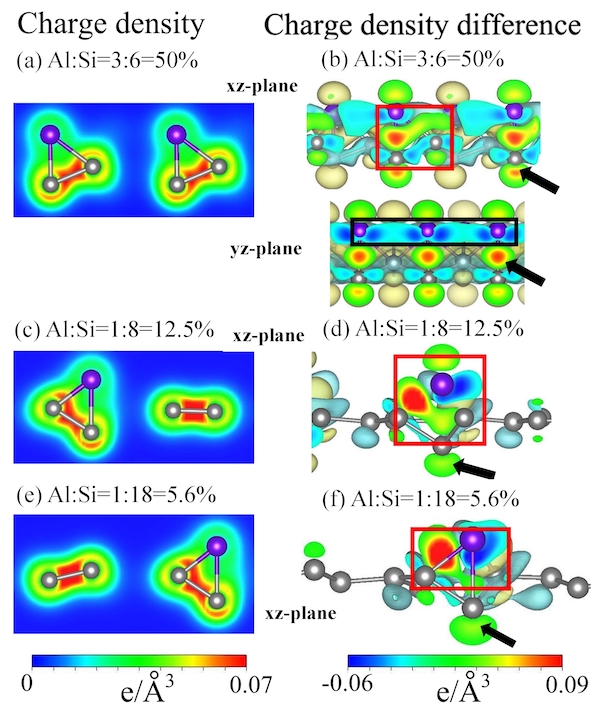}
	\caption{(Color online) The spatial charge distribution/their variations after Al-chemisorptions along the \textbf{xz}- and \textbf{yz}-planes: (a)/(b) [3:6], (c)/(d) [1:8], and (e)/(f) [1:18].}
	\label{FIG:9}
\end{figure}      

\renewcommand{\thefigure}{S1}
\begin{figure}[hp]
	\graphicspath{{figure}}
	\centering
	\includegraphics[scale=0.3]{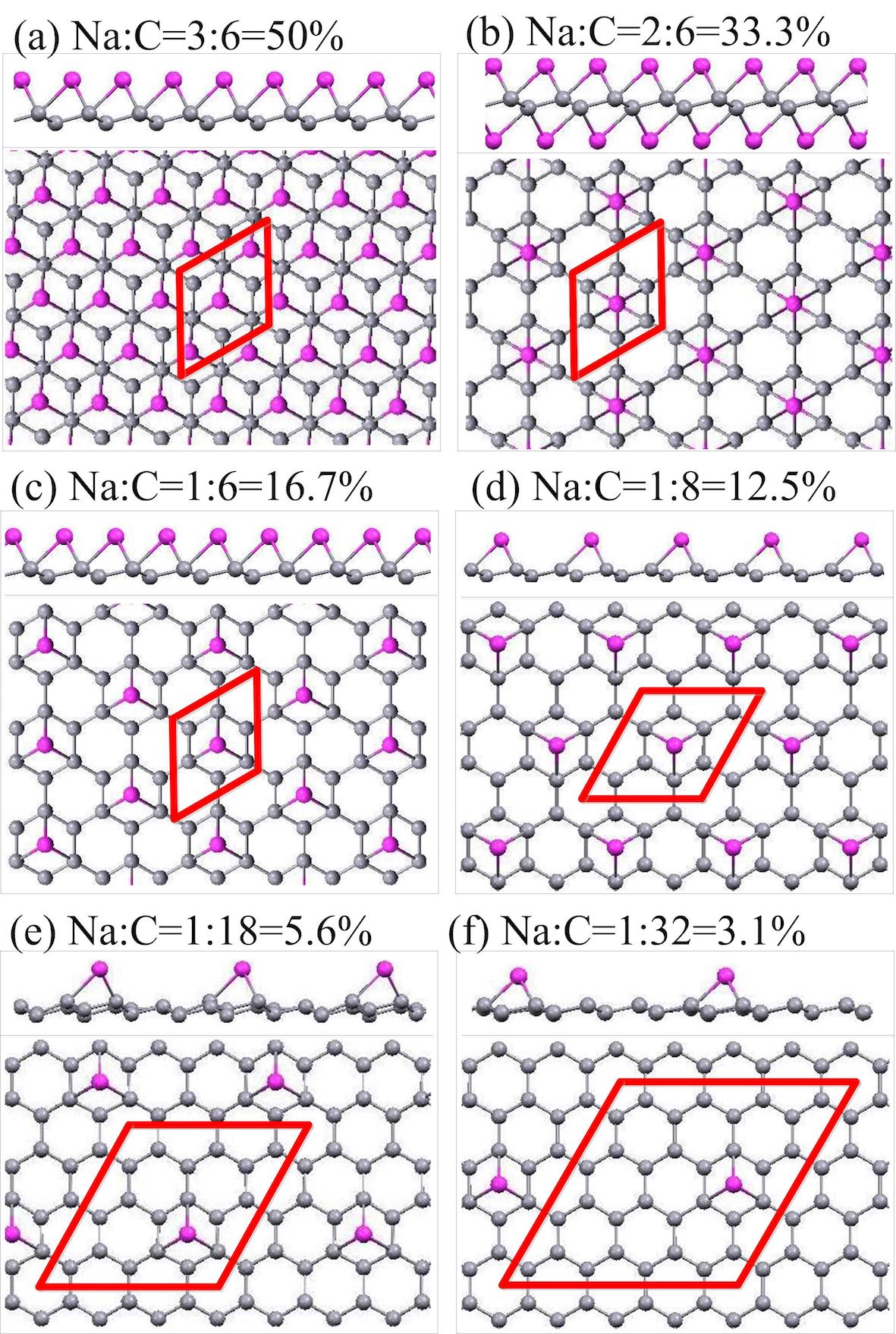}
	\caption{(Color online) The optimal geometries of the sodium-adsorbed silicenes with the side and top views for the distinct adsorptions: (a) [3:6], (b) [2:6], (c) [1:6], (d) [1:8], (e) [1:18], and (f) [1:32].}
	\label{FIG:S1}
\end{figure}      

\newpage
\renewcommand{\thefigure}{S2}
\begin{figure}[hp]
	\graphicspath{{figure}}
	\centering
	\includegraphics[scale=0.3]{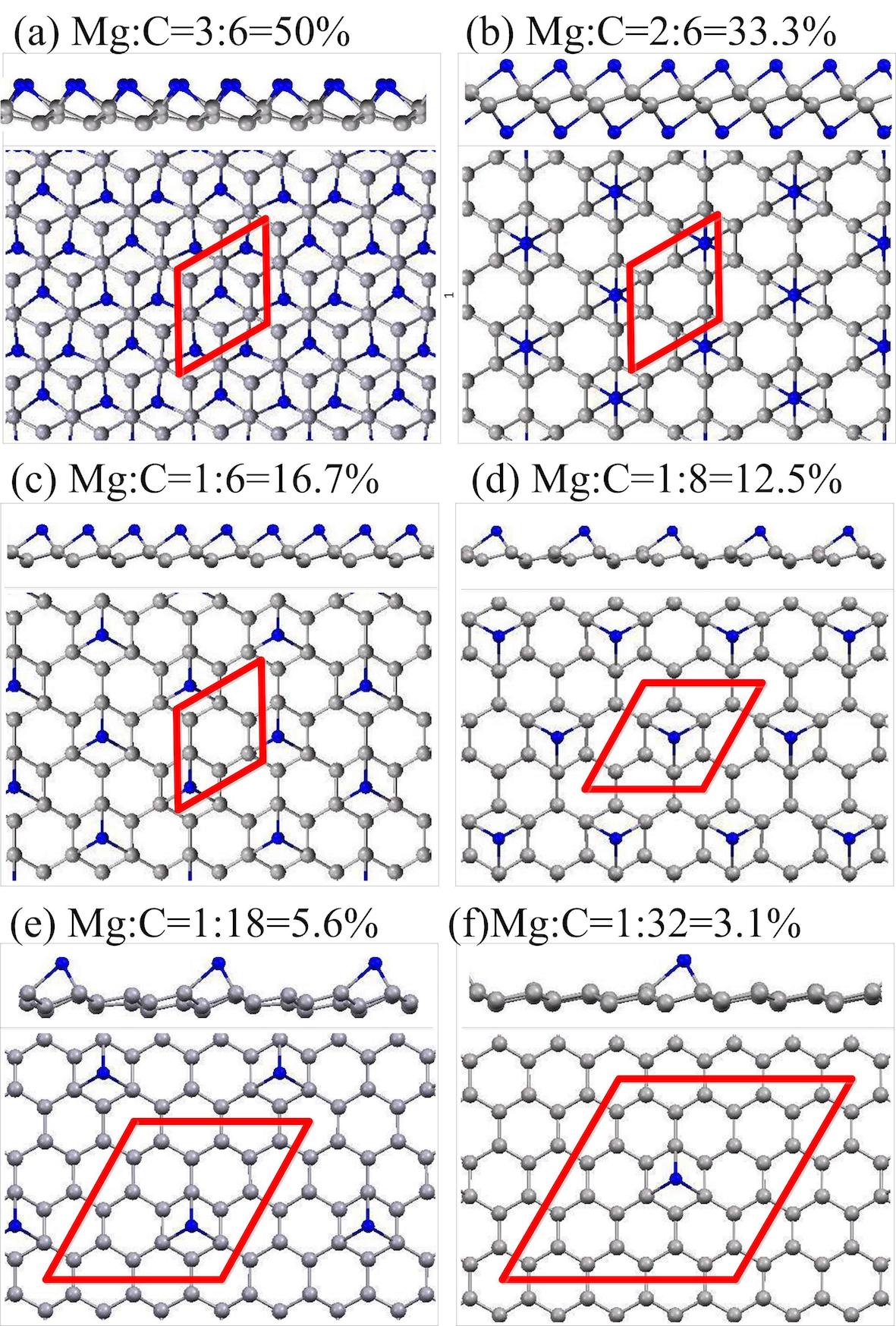}
	\caption{(Color online) The optimal geometries of the magnesium-adsorbed silicenes with the side and top views for the distinct adsorptions: (a) [3:6], (b) [2:6], (c) [1:6], (d) [1:8], (e) [1:18], and (f) [1:32].}
	\label{FIG:S2}
\end{figure}

\newpage
\renewcommand{\thefigure}{S3}
\begin{figure}[hp]
	\graphicspath{{figure}}
	\centering
	\includegraphics[scale=0.3]{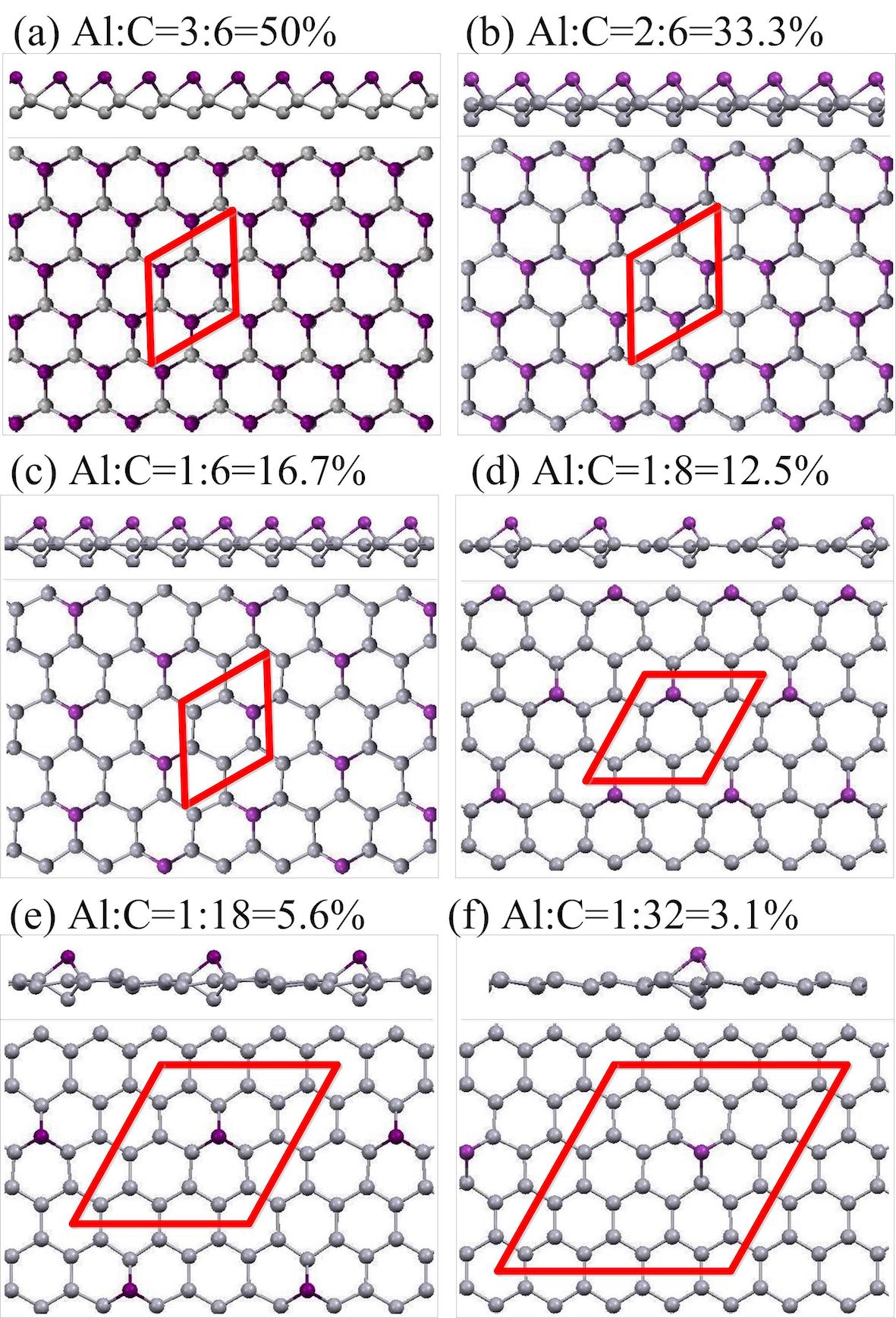}
	\caption{(Color online) The optimal geometries of the aluminum-adsorbed silicenes with the side and top views for the distinct adsorptions: (a) [3:6], (b) [2:6], (c) [1:6], (d) [1:8], (e) [1:18], and (f) [1:32].}
	\label{FIG:S3}
\end{figure}

\end{document}